\documentstyle[12pt,aaspp4]{article}

\begin{document} 

\title{ Constraints on the Physical Parameters of TeV Blazars}

\author{ Fabrizio Tavecchio and  
	Laura Maraschi \\ Osservatorio Astronomico di Brera, via Brera 28, 20100 Milano,
Italy
	\and
	Gabriele Ghisellini \\ Osservatorio Astronomico di Brera-Merate, via Bianchi 46,
22055 (Lecco), Italy. }

\begin{abstract} 

We consider the constraints that can be derived from the spectral 
shape and variability of TeV blazars on the homogeneous synchrotron 
self--Compton (SSC) model.
Assuming that the relativistic electron spectrum is a broken power law, 
where the break energy is a free parameter, we write analytical 
formulae that allow to connect observable quantities to the physical 
parameters of the model.We give 
approximate analytic formulae  also for the case of Compton scattering 
occurring
in the Klein Nishina regime, which is particularly relevant for TeV blazars.
In particular we
find that even in the latter regime a power law component can be present at the
highest energies.
 Further restrictions in the
parameter space are set assuming that the break energy results from
a balance between cooling and escape and that the soft photon lags
 measured in some sources derive from radiative cooling of high energy particles.
The constraints can be summarized as allowed regions in 
the Doppler factor -- magnetic field parameter space and are in principle
sufficient to univocally determine the model parameters and their uncertainties.

We  apply the method to three well studied sources Mkr 421, PKS 2155--304
and Mkr 501.
For Mrk 421 the available data are sufficient to fully constrain the model.
The additional restrictions are found to be consistent with the model parameters,
  supporting the proposed interpretation.
In the case of PKS 2155--304, not yet detected in the TeV band,
we estimate the peak frequency of the Compton component 
($\simeq 40$ GeV) and the expected TeV flux.  
The derived physical parameters are similar for the two sources, with 
a relatively large value of the Doppler factor ($\delta \sim 25$) 
and a magnetic field of $B\simeq 0.2$ G.
For Mrk 501 we consider both the historical low state and the flaring state
observed in April 1997. 
In the first case consistency between the various assumptions is 
reached for $\delta \simeq 10$ and $B \simeq 0.3$ G.
For the high state a similar value of the Doppler factor and a somewhat
larger value of the magnetic field are indicated, 
while the extremely large frequency of the observed synchrotron peak 
requires continuous injection/reacceleration of relativistic particles.

\end{abstract}


\section{Introduction}

 The discovery of intense gamma-ray emission from blazars had a strong impact
on our understanding of relativistic jets. Independent evidence
of relativistic beaming  was provided by the condition
that the source should be transparent to high energy gamma-rays
(Mattox et al. 1993, von Montigny et al. 1995, Dondi \& Ghisellini 1995). 
The gamma-ray fluxes and spectra,
together with data at lower frequency allowed
detailed estimates of the physical parameters of different models (Maraschi, 
Ghisellini and Celotti 1992; Blandford 1993; Sikora, Begelman and Rees 1994; 
Dermer and Schlickeiser 1993). 

More recently the discovery of TeV emission from some objects (Mrk 421, Mrk 501 and
1ES 2344+514)
and the detection of very fast variability at these high energies  
motivated new theoretical and observational studies 
(e.g. Celotti, Fabian and Rees 1998) 

The objects detected at TeV energies have similar spectral
energy distributions (SED) from radio to X-rays. They emit synchrotron radiation
with power peaking in the X-ray band.
 For Mrk 421 and Mrk 501 the X-ray and TeV radiation vary in a 
correlated fashion (Macomb et al. 1995, Buckley et al. 1996, Catanese et al. 1997,
Schubnell 1997) suggesting that the TeV emission derives 
from the same high energy electrons responsible for the synchrotron X-rays.
For Mrk 421 the TeV flux has been observed to vary with a 20 min timescale (Gaidos et
al. 1996),
which strongly limits the size of the emitting region.
Furthermore  a lag has been measured between soft and medium 
energy X-ray photons. A lag of the same order between soft and medium
X-ray photons has been measured also in PKS 2155-304, whose spectral energy
distribution closely resembles that of
 Mrk 421 up to the X-ray range, but has not yet been observed
at TeV energies due to its southern declination. The lags strongly suggest 
a radiative origin yielding a new important constraint on the physical parameters
of the emitting region.   
For Mrk 501, similar in spectral shape to the previous two, an extraordinary flaring
behaviour has been observed recently (Pian et al. 1998). 

The simplest model proposed for TeV blazars involves a single zone 
and a single population of  relativistic electrons
emitting  synchrotron radiation from radio to X-rays and inverse Compton
radiation from X-rays to gamma-rays.  
For this class of sources the seed photons for the inverse Compton process are
likely to be the synchrotron photons themselves (SSC) (see e.g. Ghisellini \&
Maraschi 1996, Mastichiadis \& Kirk 1997),
 although other models have
also been considered (for a review see Sikora 1997).
 We will limit ourselves to the 
SSC model in the following.

Our scope is  to provide general analytic expressions for all the constraints that can
be envisaged  within 
this simple model and discuss the consequent restrictions in the parameter space
 for the above mentioned sources individually and as  a group.
A similar approach was initiated by Ghisellini, Maraschi and Dondi (1996)
 (hereafter GMD), however it did
not include all the conditions and was limited to regimes well below
the Klein Nishina limit, while  for TeV blazars the Klein Nishina corrections are
important. Work along these lines has been developed by Dermer, Sturner
and Schlickeiser (1997) and by Bednarek and Protheroe (1997). 
The latter authors chose to use  precise formulas requiring however
numerical computations, with loss of generality.

The paper is organized as follows. In section 2 we discuss the constraints that can be
put on the homogeneous SSC model using all the available observations.
In section 3  the relevant formulae are derived which link
 model parameters to observative quantities. 
The effects of the Klein Nishina limit are discussed in detail and revised analytic
formulae are given valid also in the latter regime. 
In section 4 we apply our diagnostics to the case of three well known
 blazars and finally (Section 5) we discuss the results.

\section{Model constraints}

The homogeneous SSC model assumes  that radiation is produced in a single zone 
of the jet  (approximated here for simplicity as a sphere with radius $R$),
 relativistically moving at small angle $\theta $ to the observer's line of sight.
Photons up to the X-ray range  are produced 
by relativistic electrons through the
synchrotron process and are subsequently inverse Compton scattered by the same electrons
to energies in the $\gamma $-ray range.
The observed radiation  will be strongly affected by relativistic effects.
The key parameter is  the Doppler factor 
$\delta=[\Gamma (1-\beta \cos \theta )]^{-1}$, where $\Gamma $ is the bulk Lorentz
factor and $\beta =v/c$.

The observed SED 
of  TeV emitting blazars  (in a $\nu F_{\nu }$ plot) is characterized by
 two broad peaks, the first approximately
in the X-ray  band, the second in the 100 GeV-1 TeV
band . Below and above the peaks the spectrum is quite smooth and
 can be approximated with  power-law profiles with indices $\alpha _{1},\alpha _{2}$ 
respectively
where $\alpha_1 < 1$ and $\alpha_2 >1$ (see e.g. GMD). Note that within the present
 uncertainties
the "left" side of both peaks can be described by the same spectral index.
Note also that the spectrum in the TeV range is poorly known and may be affected by 
intergalactic absorption (see e.g. Stecker and De Jager 1997 and reference therein).
In figure (\ref{prima}) we show the observed SED of Mrk 421 at two different epochs
(Macomb et al. 1995). 

The observed spectral shape requires that the
the relativistic electron spectrum steepens with increasing energy. We approximate 
 this behaviour with a broken power law (GMD) 
with indices $ n_1 < 3, n_2 > 3$,  
respectively below and above the "break"energy  $\gamma _b m_e c^2$:
\begin{equation}
N(\gamma )=\left\{ \begin{array}{ll}
                    K\gamma ^{-n_1}  &  \mbox{if $\gamma < \gamma _b$} \\
		    K\gamma _b^{n_2-n_1} \gamma ^{-n_2}  &  \mbox{if $\gamma > \gamma _b$}
		   \end{array}
	   \right. 		
\end{equation}
With  these approximations we can completely specify the model using
 7 parameters: the magnetic field intensity
 $B$, $R$, $\delta $, the slopes $n_1$ and $n_2$, the
Lorentz factor of the electrons at the "break", $\gamma _b$, and the
electron density parameter $K$. The peak synchrotron power is emitted
by electrons with the break energy $E_b=\gamma _b m c^2$. The maximum energy
$\gamma _{max}$ attained
by electrons as well as a possible lower limit $\gamma _{min}$ are unimportant here
provided that $\gamma _{max} \gg \gamma _b$ and $\gamma _{min} \leq 100$

Spectra computed with the above assumptions are compared with
  the  Mkn 421 data in Fig. 1.
  The theoretical curves show that the double power-law approximation can
reproduce reasonably well the shape of the observed SED.

The available data on the SED can be used to derive 6 "observable" quantities
of particular relevance for the above model, namely
 the indices $\alpha _1$ and $\alpha _2$, the frequency of the
synchrotron and inverse Compton peaks,  
$\nu _s$ and $\nu _c$, and the peak luminosities $L_s(\nu _s)$ and $L_c(\nu _c)$.
These quantities are rather uncertain being ideally the result of 
a fitting procedure on simultaneous multifrequency data, but in practice often 
inferred from observations obtained at different epochs and with incomplete
frequency coverage.
The best sampled spectral region is usually the optical to X-ray range, therefore
$\alpha _1$, $\alpha _2$, $\nu _s$ and $L_s(\nu _s) $ are better determined than 
$\nu _c $ and $L_c(\nu _c)$. Fig. 1 can serve as an illustration of the uncertainties
in the determination of the "observables" for Mrk 421, perhaps the object
with the best overall data.
We recall that the one-zone homogeneous model is selfabsorbed at radio frequencies
and cannot explain the radio emission which implies further
contributions from the outer regions of the jet.

It is  important that, at least in principle, one can derive from the broad band
spectral observations 6 quantities that have to be 
reproduced with the 7 parameter model. One more
 observable quantity is sufficient to close the system. This can be provided
 by the minimum timescale of 
variation, $t_{var}$, which can be directly connected to the source dimension, $R$,
through the causality relation $ R \leq ct_{var}\delta $. This relation gives only an
upper limit on the dimension of the source, but a small dimension implies a high
Doppler factor if the transparency condition  is taken into account (see below).
 In order not to force $\delta$ to be too large we will
adopt limits to  the radius of the emission region in the range $ct_{var}\delta$-
$ \frac{1}{3}ct_{var}\delta$.
At this point the system is closed and any additional constraint that can be satisfied
should be regarded as supporting the model. 
These are:
\begin{itemize}
\item the transparency of the source to $\gamma $-rays. 

The high energy
photons may interact with the low energy ones producing pairs. The fact that we
observe the TeV emission poses a strong limit to the optical depth of the source
to photon photon interaction and therefore to 
the energy density of the soft radiation. From this limit one can obtain a lower limit
on the value of the Doppler factor (see e.g. Dondi \& Ghisellini 1995).  

\item the radiative interpretation of the time lags in the light curves at different
frequencies.

Recently, observations of Mrk 421 and PKS 2155-304 with ASCA (0.3-10 keV)
 have revealed that when pronounced flares occur, the soft photons lag the harder ones
 ( Takahashi et al. 1996, Urry et al. 1997).
The simplest way to explain the time lags in a homogeneous model is through the
hypothesis that they are associated to the time necessary
 for freshly injected high energy  electrons to cool.
 With this assumption the  measured time lags determine
  the cooling time
setting a well defined condition on the model. 

\item the consistency of the break energy of the electrons with the condition of
balance between cooling and escape.

In our discussion the electron energy distribution is not calculated
self-consistently (for an example of this alternative approach see e.g. Inoue and 
Takahara 1996, Mastichiadis and Kirk 1997, Ghisellini et al. 1998). The 
two spectral indices and the
break energy are determined from the observed spectra.
 Having determined at the same
time the dimension of the emission region and the magnetic and radiative 
energy densities we can verify a posteriori whether the cooling time of the
electrons at the break
is of the order of the escape (expansion) time from the emission region.
We will assume that consistency
is achieved if the required escape/expansion velocity is in the range $c/3-c$.
We recall however that in the standard hypothesis of a power law injection and
subsequent cooling and escape in a homogeneous region, a spectral change
$\Delta \alpha = 0.5$ is expected, while in general for the sources discussed below 
the observed steepening is larger, suggesting a more complex situation.

We approximate the cooling
time  as the minimum between the synchrotron and the
Compton cooling times. In the $B$--$\delta $ parameter space the regions of validity
of these two approximations are separated by the line expressing equipartition
 between the magnetic and radiation energy density. In each of these regions 
a value of the cooling time defines only one line.
 Bednarek \& Protheroe (1997) discuss a similar constraint on the cooling time
 of the electrons, 
but they incorrectly apply the condition to both  the synchrotron and the
Compton cooling times {\it simultaneously} obtaining
 two independent constraints. 
\end{itemize}   
Summarizing, the SSC homogeneous model needs 7 independent parameters: the spectral
observations give 6 independent quantities, and the seventh observational
information used
to close the system is the minimum variability timescale.
Other constraints are provided
 by the optical depth for $\gamma $-$\gamma $ interaction, by the observed time 
lags between soft and hard X-ray photons
and by the consistency of the value of $\gamma _b$. These last conditions are
additional independent constraints: if verified, these  constraints argue
for the physical consistency of a homogeneous 
SSC model.\\
 
\section{Relations between the model parameters and observables}  

In this section we derive the fundamental relations between the model parameters 
and the observable quantities discussed above.
 We chose to do this analytically at the cost of
some inevitable approximation in order to maintain generality. Most of these 
formulae are also discussed in GMD and Dermer et al. (1997). 
 
We assume that the two spectral indices can be determined directly from the data,
independently of other quantities.
The structure of the relations is then such that we can eliminate 3 of the
remaining 5 parameters
(i.e. $\gamma_b$, K and R) and express the constraints as allowed regions in the 
B-$\delta$ parameter space.

\subsection{Peak frequencies}

In the scheme that we consider the dominant synchrotron power is emitted by
 electrons at the break (those with Lorentz factor 
$\gamma _b$). Primed quantities refer to the blob frame so that the observed frequency
is given by $\nu _s=\nu _s^{\prime }\delta$.
The photons at $\nu _s$ dominate the energy density of
the target photons and  the inverse Compton power in the Thomson limit will be 
dominated by photons at
frequency $\nu _c=(4/3)\gamma _b^2\nu _s^{\prime }\delta$, produced in the scattering
between the electrons at the break and the photons with frequency $\nu _s^{\prime }$
(see also GMD).
Inverting the last equation one obtains:

\begin{equation}
\gamma _b=\left( \frac {3\nu _c}{4\nu _s} \right) ^{1/2}
\label{gb}
\end{equation}

\noindent
Therefore the ratio between the peak frequencies gives directly 
the value of $\gamma _b$.\\

\noindent
The synchrotron frequency averaged over the spectral shape 
for an electron of Lorentz factor $\gamma_b$ is: 

\begin{equation}
\nu _s=3.7\cdot 10^6\gamma _b^2B \frac{\delta }{1+z}
\label{nus}
\end{equation}

\noindent
using the value of $\gamma _b$ given by eq. (\ref{gb}), gives: 

\begin{equation}     
B\delta =(1+z)\frac{\nu _s^2}{2.8\cdot 10^6 \nu _c}
\label{cinque}
\end{equation}

\noindent
The last formula shows that for fixed $\nu _s$ and $\nu _c$, $B$ and $\delta $
are inversely proportional.
We will see in the next section that in the KN limit this situation is reversed.

\subsection{Peak luminosities}

Another relation can be obtained from the ratio of the total luminosity of the 
synchrotron peak and the total luminosity of the self-Compton peak, which is directly
related to the ratio between the radiation and the magnetic field
energy densities inside the source (in the comoving frame):

\begin{equation}
\frac{L_c}{L_s}=\frac{U^{\prime }_{syn}}{U^{\prime }_B}
\label{natan}
\end{equation}

\noindent
For our aims it is useful to express the total luminosity through
 the peak luminosity, which is more accessible
to observations. 
A simple calculation shows that  the total luminosities
($L_s$ and $L_c$) and the peak luminosities [$L_s(\nu _s)$ and $L_c(\nu _c$) ]
are related by:

\begin{equation}
L_{s,c}=f(\alpha _1,\alpha _2)\nu _{s,c}L_{s,c}(\nu _{s,c})
\label{royan}
\end{equation}

\noindent
where $f(\alpha _1,\alpha _2)$ is given by

\begin{equation}
f(\alpha _1,\alpha _2)=\frac{1}{1-\alpha _1} + \frac{1}{\alpha _2-1}
\end{equation}

\noindent
Expressing $U^{\prime }_{syn}$ as: 

\begin{equation}
U^{\prime }_{syn}=\frac{L_s}{4\pi R^2c\delta ^4}
\label{density}
\end{equation}

\noindent
we can rewrite eq. (\ref{natan}) as:

\begin{equation}
\frac{L_c}{L_s}=\frac{\nu _cL_c(\nu _c)}{\nu _sL_s(\nu _s)}=\frac{2\nu _sL_s(\nu _s)f(\alpha _1,
\alpha _2)}{R^2c\delta ^4B^2}
\end{equation}

\noindent
If we express the source dimension $R$ with the variability timescale

\begin{equation}
R\leq ct_{var}\delta (1+z)^{-1}
\label{undici}
\end{equation}

\noindent
we can finally write:

\begin{equation}
B \delta ^{3} \geq (1+z) \left[ \frac{ 2(\nu _sL_s(\nu _s))^2f(\alpha _1,\alpha _2)} 
{c^3t^2_{var} \nu _cL_c(\nu _c)} \right]^{1/2}
\label{dodici}
\end{equation}

\noindent
This formula relates the magnetic field intensity and the Doppler factor to
observable quantities.
Eq. (\ref{dodici}) is  an inequality because of the use of the inequality
eq. (\ref{undici}); taking $ct_{var}/3$ as a lower limit on the radius of the source,
as discussed in Section 2, 
eq. (\ref{dodici}) will be represented in the $B$, $\delta $ space by a strip of 
finite width.
Thus the intersection of eq. (\ref{dodici}) with  eq. (\ref{cinque}) allows  
to estimate $B$ and $\delta$.

\section{ The Klein Nishina regime}

\subsection{Spectral shape of the Inverse Compton emission}

The adopted spectral shape for the electrons ensures that 
 the synchrotron spectrum has a well defined peak (break) (as observed). Thus
both, electrons and photons, have a low energy branch and a high energy branch
(below and above the break). With the adopted approximations one
can compute the inverse Compton spectrum as the sum of 4 separate integrals:
 $I_{1,1}, I_{1,2}$ and $ I_{2,1}$, $I_{2,2}$, representing the contributions
from electrons of the low (1) and high  (2) energy  branches scattering
 photons of the low and high energy branches respectively. Full expressions are
given in the Appendix. This subdivision is very useful
 in the discussion of the Klein Nishina limit.

The different contributions, computed 
 using the well known $\delta $-function approximation for the single electron
emission and a step function for the energy dependence of the cross section, 
$\sigma =\sigma_T$
for $\gamma \nu ^{\prime } < 3/4$ and  $\sigma =0$ otherwise 
(see  Coppi and Blandford 1990 for a critical discussion), are shown in Fig. 2.
All the spectra are computed in the source frame for fixed values 
of $\nu ^{\prime }_s=10^{15}$ Hz and varying $\gamma _b$ from $10^3$ to $10^6$,
that is for increasing importance of the KN limit. 

Since the assumed particle spectra have no high energy cut offs, at high enough
energy, the KN limit is always relevant and steepens the spectrum. The limiting
frequency can be estimated considering that the largest contribution to the 
radiation energy density is due to photons of frequency $\nu ^{\prime }_s$.
Therefore the KN limit will set in at a  frequency 
$\nu ^{\prime } _K \geq (mc^2/h)^2/\nu ^{\prime }_s$.
This is seen in Fig. 2a and 2b, where
three (12, 21, 22) of the four components in the total spectrum
cut off at high energies, while the peak of the inverse Compton spectrum
occurs as expected in the Thomson limit. 

However the effect will be severe and affect 
both  the value of the peak frequency and the bulk of the emitted power 
if  $\nu_K$ is close to $\nu_C$,
that is if the KN regime sets in already for the scattering of electrons of energy
$\gamma _b$ with photons at the peak of the synchrotron emission. 
Therefore the condition for the KN regime to affect severely the luminosity and the peak
frequency of the inverse Compton emission, requiring a revision of the formulae
given in section 3 is:

\begin{equation}
\gamma _b\nu _s^{\prime } \geq \frac{3}{4} \frac{mc^2}{h}
\label{condizione}
\end{equation}

\noindent
Using eq. (\ref{gb}) for $\gamma _b$ to express the above condition
with observed quantities, eq. (\ref{condizione}) becomes a condition on $\delta $:

\begin{equation}
\delta < \delta _{KN}=\left[ \frac{\nu _c\nu _s}{(3/4)(mc^2/h)^2} \right]^{1/2}
\label{limite}
\end{equation}

Since eq. (\ref{gb}) (valid in the Thomson limit) gives the lowest possible value of
$\gamma _b$, eq. (\ref{limite}) gives the limit above which the full KN regime sets
in. However even for $\delta $ somewhat above the KN limit corrections are non
negligible (see Section 4.2)

The contribution involving both electrons and photons on the high
energy branches ( $I_{2,2}$ ) is already strongly suppressed in Fig. 2b and disappears 
completely when the KN condition is met (Fig. 2c, 2d). 
 The other contributions are reduced according to the
availability  of photons below the KN threshold. 
 In Fig. 2c and 2d $I_{1,2}$ is negligible and the low energy side is dominated by 
 by the $I_{1,1}$ branch. 

It is interesting to note that for the assumed spectral shape
 the scattering of highest energy electrons with very low energy photons 
gives a significant power law contribution to the inverse Compton
spectrum at the highest energies. The latter component is not important energetically
(at the peak) however it provides a high energy tail which extends the range 
of the inverse Compton emission. 
The spectral index of the latter contribution can be easily derived from the
expression for $I_{2,1}$ given in the Appendix and is $\alpha _{KN}=2\alpha _2 
-\alpha _1$. This component could explain the  power law spectra measured in the
 TeV range for Mrk 421 and Mrk 501 (Zweerink et al. 1997).

\subsection{Compton peak frequency and luminosity in the KN limit}

In the KN regime both the expression of the Compton peak frequency and luminosity must
be modified.
As shown in Fig. 2c, 2d, with the assumed electron spectrum  the peak 
frequency in the KN
limit is determined by the scattering between all the electrons and 
photons with $\nu <\nu _s$. 
Maximizing the sum of $I_{1,1}$ and $I_{2,1}$ (see Appendix) leads to the following
relation between $\nu _c$ and $\gamma _b$:

\begin{equation}
\nu _c = \nu ^{\prime }_c \frac{\delta }{1+z} \simeq \frac {mc^2}{h} 
\gamma _b g(\alpha _1,\alpha _2 )\frac{\delta }{1+z} 
\label{intef}
\end{equation}
where:
\begin{equation}
g(\alpha _1,\alpha _2)=\exp \left[ \frac{1}{\alpha _1 -1} +\frac{1}{2(\alpha _2-
\alpha _1)} \right]
\end{equation}
is a factor smaller than one.  

\noindent
Substituting $\gamma _b$ derived from eq. (\ref{intef}) in eq. (\ref{nus}) we
obtain:

\begin{equation}
\frac{B}{\delta }=\frac {\nu _s}{\nu _c^2} \left( \frac {mc^2}{h} \right)^2 \frac {g(\alpha
_1, \alpha _2)^2 }{3.7\cdot 10^6} \frac{1}{1+z}
\label{cinquekn}
\end{equation}

\noindent
which is the analogous of the eq. (\ref{cinque}) for the KN regime. Note that in this equation
$B$ and $\delta $ are directly proportional.

Having obtained a revised expression of eq. (\ref{gb}) we can use the latter (eq.
[\ref{intef}]) to express (\ref{condizione}) through observables. We therefore arrive
at a revised version of the threshold for the KN regime:

\begin{equation}
\delta < \delta _{KN}=\left[ \frac{\nu _c\nu _s}{(3/4)(mc^2/h)^2} \right]^{1/2}
g(\alpha _1,\alpha _2)^{-1/2}
\label{limitedue}
\end{equation}

Since $g(\alpha _1,\alpha _2)$ is less than 1, eq. (\ref{limitedue}) is more stringent
than eq. (\ref{limite}): therefore we find two 
{\it
different} limits dividing the regions of applicability of the two treatments. Between
this two limits there is a transition region, where both the approaches are
approximations.  

In the KN regime the power in the Compton peak is substantially
diminished.    
equation (\ref{natan})  should  be replaced by:

\begin{equation}
\frac{L_c}{L_s}=\frac{U^{\prime }_{syn, avail}}{U^{\prime }_B}
\label{taita}
\end{equation}

\noindent
where $U^{\prime }_{syn, avail}$ includes only photons  below the KN limit
for electrons with the break energy, that is with
$\nu ^{\prime }_0 \leq mc^2/{h \gamma _b}$. We can write:

\begin{equation}
U^{\prime }_{syn, avail}=\int _{0}^{3mc^2/4h \gamma _b} \epsilon ^{\prime }_{syn}(\nu ^{\prime
}_0) d\nu ^{\prime }_0 
\end{equation}

\noindent
where $\epsilon ^{\prime }_{syn}(\nu ^{\prime }_0)$ 
is the energy density of the synchrotron radiation for a given frequency.
A simple calculation shows that the integral is given by:

\begin{equation}
U^{\prime }_{syn, avail}=U^{\prime }_{syn} \left( \frac {3mc^2\delta }{4 h \gamma _b \nu
_s} \right)^{1-\alpha _1}
\label{ild}
\end{equation}

\noindent
where the factor in parentheses represents the ratio between the available energy
density and the total energy density. From the last relation we can obtain the
following expression for the energy density ratio in the comoving frame in the KN regime:

\begin{equation}
\frac{U^{\prime }_{syn}}{U^{\prime }_B}=\frac{L_c}{L_s} \left[ \frac{3}{4}
\left(\frac{mc^2}{h} \right)^2
\frac{\delta ^2}{\nu _s\nu _c} \frac{g(\alpha _1,\alpha _2)}{1+z} \right]^{\alpha _1-1}
\end{equation}

\noindent 
Using the last formula and expressing the total luminosities
through eq. (\ref{royan}) and the radius with eq. (\ref{undici}) we can finally obtain 
the analogous of eq. (\ref{dodici}) for the KN limit:

\begin{equation}
B\delta^{2+\alpha _1} > (1+z)^{\alpha _1} \left[ \frac{g(\alpha _1,\alpha _2)} {\nu
_c\nu _s} \right] ^{(1-\alpha _1)/2} 
\left[ \frac {2 (\nu _sL_s(\nu _s))^2 f(\alpha _1,\alpha _2)}
{ c^3 t_{var}^2 \nu _c L_c(\nu _c)} \right]^{1/2} \left( \frac{3mc^2}{4h}
\right)^{1-\alpha _1} 
\end{equation}

\section{Additional constraints on the model parameters}

\subsection{The pair production opacity}

Another strong constraint is obtained from the condition of transparency of $\gamma
$-rays to pair-production
absorption. Following Dondi \& Ghisellini (1995) we can write this condition in the
form:

\begin{equation}
\delta > \left[ \frac{\sigma _T}{5hc^2}d_L^2(1+z)^{2\beta } \frac {F(\nu
_0)}{t_{var}} \right]^{1/(4+2\beta )}
\label{diciotto}
\end{equation}

\noindent
where $\nu _0=1.6\cdot 10^{40}/\nu _{\gamma }$ is the frequency of target photons 
and $\beta $ is the spectral index of the target photons
($\alpha _1$ for $\nu _0<\nu _s$ and $\alpha _2$ for $\nu _0>\nu _s$).
The transparency condition does not depend on the specific
emission mechanisms: therefore it provides a strong
independent constraint on the minimum value of the Doppler 
factor in the SSC model.

\subsection{Cooling time}

The radiative cooling time for a synchrotron--self Compton emitting electron is given
by:

\begin{equation}
t^{\prime }_{cool}=\left[ \frac {4}{3} \frac{\sigma _Tc}{mc^2} \gamma (U^{\prime }_B 
+U^{\prime }_{syn, avail}) \right]^{-1}
\end{equation}

\noindent

Clearly the cooling time is determined by the fastest of the two cooling processes
that is  synchrotron cooling for $ U^{\prime }_B > 
U^{\prime }_{syn, avail} $ or inverse Compton cooling for 
$ U^{\prime }_B < U^{\prime }_{syn, avail} $. 
Let us first consider for simplicity the Thomson regime. 
In the case in which the magnetic field energy density dominate we can approximate the 
cooling time as (after some manipulations):

\begin{equation}
t^{\prime }_{cool} \simeq t^{\prime }_{sync} = C_s B^{-3/2} \delta ^{1/2} \nu ^{-1/2}
\end{equation}

\noindent
while in the opposite case ( using eq. [\ref{density}] for the energy density and 
expressing $\gamma $ with eq. [\ref{nus}] ): 

\begin{equation}
t^{\prime }_{cool} \simeq t^{\prime }_{comp}= C_c B^{1/2} \delta ^{13/2} \nu ^{-1/2}
\end{equation}

\noindent
where $C_s$ and $C_c$ are constants.

In the observer frame:

\begin{equation}
t_{sync} = C_s B^{-3/2} \delta ^{-1/2} \nu ^{-1/2}_s ; \ t_{comp}= C_c 
B^{1/2} \delta ^{11/2} \nu ^{-1/2}_s
\end{equation}
The equipartition condition  $ U^{\prime }_B = 
U^{\prime }_{rad} $ divides the regions of validity of the two approximations
in the $B-\delta $ plane.  The Compton cooling approximation will be valid
 in the lower region (where $ U^{\prime}_B < U^{\prime}_{sync}$), while the
 synchrotron cooling approximation will
be valid in the upper region. In both cases a given value of $t_{cool}$ defines 
an inverse relation between B and $\delta $, that is lines with negative slopes,
intersecting each other at the equipartition line (see Figures in the Application 
sections). Thus a condition on the cooling time corresponds to a minimum value of 
$\delta$ above which the condition is represented by a double valued
function with two branches. Lower values of $\delta$ correspond to shorter cooling
times.

\subsection{Time lags}

The measurement of time delays between variations in different energy bands 
provides an important  restriction on the possible values of the physical 
parameters of the model.
The most natural way to explain the lags in a homogeneous model is to interpret
them as due to the cooling time of the emitting electrons (Takahashi et al. 1996, 
Urry et al. 1997). In this picture the quiescent emission is produced by the
"stationary" population of electrons, 
while the flare is due to the injection of a monoenergetic
population of high energy electrons:
while these electrons cool their emission drifts to lower frequencies as observed.
Adopting this hypothesis we can relate the observed time lags to the 
cooling time
of the electrons and therefore to the physical parameter of the source.
If the cooling is dominated by the synchrotron emission (i.e. $ U_B ^{\prime }\geq 
U_{syn}^{\prime } $,
where $ U_{syn}^{\prime } $ is $U_{syn,avail}^{\prime } $ in the KN limit), we  
obtain the following relation (see also Takahashi et al. 1996):

\begin{equation}
B\delta^{1/3}=300 \left( \frac{1+z}{\nu _{1,17}} \right)^{1/3}  
\left[ \frac{1-(\nu _1/ \nu _0)^{1/2}}{\tau _{obs}} \right]^{2/3} G
\label{diciassette}
\end{equation}

\noindent
where $ \nu _0$ and $ \nu _1$ are the X-ray frequencies ( $\nu _0 > \nu _1$) and $\tau _{obs}$ is the 
measured lag (in seconds).
In the Compton cooling dominated region ( i.e. $ U_B ^{\prime }\leq U_{syn}^{\prime }$)
we can write the equivalent relation:

\begin{equation}
B\delta ^{11}=\frac {3.7\cdot10^{-8}}{1+z} \nu _{1,17}\left[ \frac {2\nu _sL_s(\nu _s)
f(\alpha _1, \alpha _2) }
{c^3t_{var}^2} \frac{\tau _{obs}} 
{1-(\nu _1/ \nu _0)^{1/2} } \right]^2
\label{diciassettebis}
\end{equation}

\subsection{Cooling vs. escape}

In our approach the value of $\gamma _b$ enters as a free parameter in the model and
is determined from the observations.
However in more sophisticated models $\gamma _b$
could be determined selfconsistently by the equilibrium
between injection, cooling and escape of electrons from the
source. Here we include a posteriori the condition that the cooling time at 
$\gamma _b$ is equal to the escape time.
The latter constraint can be written as:
 
\begin{equation}
\frac{R}{\beta _{esc}c} = \left[ \frac {4}{3} \frac{\sigma _Tc}{mc^2} \gamma _b(U^{\prime }_B+U^{\prime }
_{syn, avail}) \right]^{-1}= \frac{5\cdot 10^{8}}{\gamma _bB^2(1+U^{\prime }_{syn, avail}/U^{
\prime }_B)}
\label{bond}
\end{equation}

\noindent
where $\beta _{esc}$ is the electron escape velocity in unit of $c$ and $U^{\prime }_B$ and 
$U^{\prime }_{syn, avail}$ are,
respectively, the magnetic field energy density and the radiation energy density.\\ 
If cooling is dominated by the synchrotron process, expressing (\ref{bond}) in terms
of the observational quantities $\nu _s$ and $\nu _c$ gives:

\begin{equation}
B\delta^{1/2} >\left[ \frac {5\cdot10^8(\nu _s/\nu _c)^{1/2}(1+z)}{t_{var}}
\right]^{1/2} \beta _{esc} ^{1/2}
\label{sedici}
\end{equation}

\noindent
for the Thomson limit and:

\begin{equation}
B> \left[ \frac{5\cdot 10^8 g(\alpha _1, \alpha _2)}{t_{var}}\frac{mc^2}
{h} \frac{1}{\nu _c} \right]^{1/2} \beta _{esc}^{1/2}
\label{sedicikn}
\end{equation}
\noindent
for the KN limit.

\noindent
Even when Compton cooling is less important than synchrotron cooling, if it occurs in
the KN regime, the value of $\gamma _b$ must be taken from eq. (\ref{intef}) instead 
of eq. (\ref{gb}), which
causes the difference between (\ref{sedici}) and (\ref{sedicikn}).

In the Compton-cooling dominated region the equations are: 

\begin{equation}
B\delta ^{11}> 5.2\cdot 10^{-14} \left[ \frac{\nu _sL_s(\nu _s)f(\alpha _1, \alpha _2)}{c^3} 
\right]^{2.1} \left[ \beta _{esc}(\nu _s/ \nu _c)^{1/2}(1+z) \right]^{-8/5}
t_{var}^{-13/5}
\end{equation}

\noindent
for the Thomson regime and:

\begin{equation}
B\delta ^{6/(1-\alpha _1)} = \xi (\alpha _1,\alpha _2 ) \left[ \frac{\nu _sL_s(\nu _s)
f(\alpha _1, \alpha _2)}{t_{var}\beta _{esc}} \right]^{1/(1-\alpha _1)} \left[ \nu _c
(1+z)\right]^{(3\alpha _1-2)/(1-\alpha _1)}
\end{equation}

\noindent
[$\xi (\alpha _1,\alpha _2 )$ is a constant given in the Appendix] for the KN regime.

The electron escape time is not a well known
parameter and depends on the transport processes in the emitting region: in the
applications we check the consistency of the observed value of $\gamma _b$ for values
of $\beta _{esc}$ in the range $1-1/3$.

\section{Applications}

In the previous section we have shown that the constraints considered lead to
equations that can be
expressed in terms of two parameters: the magnetic field and the Doppler factor.
Each constraint then corresponds to an allowed region or to a line in the 
 Log $B$ - Log $\delta $ plane.
Three equations do not apper explicitly: one determines $\gamma_b$ from the peak
frequencies, the second determines
the radius from the variability timescale, the third the  normalization constant
 of the electron spectrum from the observed synchrotron luminosity. 
We examine here what consequences can be derived comparing  the constraints
available for the three brightest and best observed BL Lacs with synchrotron peaks
near or within the X-ray band.

\subsection{Mrk 421} 

We start from the following values for the observable quantities, derived from the
simultaneous observations of Macomb et al. (1995)
(the corresponding data are shown in Fig. 1)
: $\nu _s=3 \cdot 10^{16}$ Hz, $\nu
_c=6.5\cdot 10^{25}$ Hz, $\alpha _1=0.5$, $\alpha _2=1.75$ 
$\nu _sL_s(\nu _s)=4.7\cdot 10^{44}$
erg s$^{-1}$ and $\nu _cL_c(\nu _c)=6.5\cdot 10^{44}$ erg s$^{-1}$ (see Table 1).
These refer to the low state (for the high state we find similar results for the values
 of $B$ and
$\delta$). The TeV observations show extremely short variability timescales
 (Gaidos et al. 1996 report a flare with $t_{var}\simeq 20$ min),
while in the X-ray band variability timescales of several hours are observed (see
Takahashi et al. 1996): in Fig. (\ref{2}) we have used $t_{var}\simeq 1$ h. 
For Mrk 421 the soft photon lag within the ASCA band was measured by Takahashi et al
 (1996) yielding $\tau _{obs}=3200$ s between  $E_1=1$ keV and $E_2=5$ keV.

In Fig. (\ref{2}) we show the parameter space for Mrk 421. The 
 transparency condition (\ref{diciotto})
calculated for photons with $\nu=10^{26}$ Hz yields a lower limit $\delta >15$
represented by the vertical line.
 The KN condition  
[eq. (\ref{limite})] yields an upper limit $\delta <14$,
while the more stringent version [eq. (\ref{limitedue})]  yields $\delta <31$.
 For $14 <\delta <31$ we are therefore in
 an intermediate region where both limits are inaccurate. We chose to use the
Thompson limit here since the resulting  value of $\delta $ will be 25, close to the
upper end. We recall however that even if the Thomson limit is adequate for 
the peak of the Compton component, KN effects cannot be neglected in the computation
of spectra at higher energies (see Fig. 1, Fig. 2a). 
 
The observed ratio between the two peak frequencies [eq.
(\ref{cinque})] imposes an inverse proportionality relation between B and
$\delta $. This is shown as a shaded area allowing for a factor 3 uncertainty
in the actual value of the Compton peak frequency.

The ratio of the peak luminosities (\ref{dodici}) yields a steeper relation 
between B and $\delta$, $B \propto \delta^{-3}$. Again it is represented by 
a shaded area to allow for uncertainties in the observational  input data.

The intersection between the two regions above defines the allowed region, 
where the actual values of the parameters should fall.  This region could shrink
if the observational data improve, but given the present gap between the
energy ranges of EGRET and the ground based Cherenkov telescopes it is unlikely
that it could be substantially reduced.

The time lag constraint has a shallow dependence on $\delta $ ( $B\propto\delta^{-1/3}$)
and is well defined observationally. Therefore it is represented as
single thick line (labelled {\it l} ) in Fig. 3. 
It is extremely interesting that this line, which is completely independent of the 
spectral constraints, is consistent with the parameter region allowed by those
constraints.
It is also interesting that the region defined by the condition that the break energy
of the electrons is determined by a balance between cooling and escape
 eq. (\ref{sedici}) (region C)
encompasses the lag constraint in the relevant parameter interval. 

We conclude that for Mrk 421 the spectral constraints are consistent both with the 
radiative interpretation of the soft photon lag and with the break energy as due
to cooling. These are persuasive arguments in support of the simple general model. 
The estimated parameter values are $B\simeq 0.25$ and $\delta \simeq 25$ (see Table 2)

Note that the value of $\delta $ is rather high compared to other estimates (Ghisellini
et al. 1998) but similar to that found from direct model fits by Mastichiadis \& Kirk
(1997) and Ghisellini \& Maraschi (1996).

\subsection{PKS 2155-304}

This BL Lac object has been studied for a long time with numerous multifrequency
campaigns (Urry et al. 1997; Edelson et al. 1995). It
has been detected as a $\gamma $-ray source by EGRET on one occasion (Vestrand et
al. 1995) and more recently in a flaring state (IAU Circ. 6774). No simultaneous 
X-ray and gamma-ray observations are available yet.
It has not been detected in the TeV band. This may however be due to its southern
declination, since in the southern hemisphere  Cherenkov telescopes have started
observations only recently.

The  parameters of the synchrotron component are inferred
 from the spectrum measured during the 1994 multifrequency  
campaign ($\nu _s=10^{16}$ Hz and $\nu _sL(\nu _s)= 1.5
\cdot 10^{46}$ erg s$^{-1}$, Urry et al. 1997). Comparison with previous observations
 (see Sambruna et al. 1994) shows that
during these observations PKS 2155-304 was in an intermediate state, a factor 2 less
luminous than the brightest state recorded by EXOSAT.

The gamma-ray spectrum measured with EGRET is hard, implying that the Compton peak
is beyond the EGRET range. However the absence of TeV observations does not allow to
determine the position of  the peak. 
We will therefore use the other constraints to estimate the peak frequency
of the Compton emission and the associated TeV luminosity,
  assuming that  all of them
 should be satisfied as is the case for Mkn 421.

In Fig. (\ref{3}) we show the parameter obtained for $\nu _c = 10^{25}$
Hz and for an "observed" luminosity of $\nu _cL(\nu _c)= 1.5 \cdot 10^{46}$ 
erg s$^{-1}$ extrapolated from the GeV range.  
We assume a minumum variability time of $t_{var}=2\cdot 10^4$ s and a delay of
$\tau _{obs} \simeq 5\cdot 10^3$s between the variations observed in the 0.75 keV and 6
keV energy bands (Urry et al. 1997). 
The vertical line representing the transparency condition is
plotted for photons with frequency $\nu \simeq 10^{25}$ Hz for consistency 
with the following estimates of the TeV flux. Note, however, that the lower limit on the Doppler factor given by the
transparency condition would be weaker in the absence of TeV emission.

In this case, the largest uncertainty is on the position of the Compton
peak, while the frequency of the synchrotron peak is reasonably well known.
We therefore plot the uncertainty band for fixed $\nu _s$ and for
$\nu _c$ ranging from $2.5 \cdot 10^{24}$ Hz to $2.5 \cdot 10^{25}$ Hz.
For these values of the peak frequencies the KN limit yields $\delta _{KN} < 3$,
 smaller than the transparency limit: we can make use of the Thomson formulation.

We find that all the constraints are satisfied for values of $B$ and $\delta $ very
similar to those found in Mrk 421 ($B\simeq 0.25$ and $\delta \simeq 25$); the
cooling-escape consistency for $\gamma _b$ is satisfied also in this case.

We can estimate the value of the TeV flux of PKS 2155-304 with the hypothesis that 
beyond the break the spectral indices of Mrk 421 and PKS 2155-304 are similar. 
From the spectrum of Mrk 421 we can obtain the value $\alpha _{TeV} \simeq 2.2$; from
the inferred value of the $\gamma $ peak flux of PKS 2155-304  we can
evaluate the TeV flux as:
\begin{equation}
F_{TeV}=F(\nu _c) \left( \frac{\nu _{TeV}}{\nu _c} \right)^{-\alpha _{TeV}}
\end{equation}
\noindent
that gives an integrated flux $F(>1TeV) \simeq 10^{-11}$ ph s$^{-1}$ cm$^{-2}$, about 
 $25\%$ that of Mrk 421 in the in the non-flaring state. 
Hence we expect that a
detection  of PKS 2155-304 should be possible but requires sufficient sensitivity, 
due to the low flux.
TeV observations of PKS 2155-304 would be extremely important also because
they may reveal or constrain absorption of high energy photons by the intergalactic IR
background (Stecker and De Jager 1997), better than 
in the case of Mrk 421 because of the larger  distance.

\subsection{Mrk 501} 

The BL Lac object Mrk 501 (z=0.034) underwent a dramatic TeV and X-ray outburst
in April 1997 (see e.g. Pian et al. 1998, Catanese et al. 1997).
Previous observations showed a spectral distribution very similar to Mrk 421,
with a synchrotron peak in the soft X-ray region; in contrast, during the April
 observation the peak was found at about 100 keV, a shift in energy of at least a
factor 100. The broad band spectral shape leads to
interpret the hard X-ray emission as
synchrotron radiation and the flare as due to injection of electrons
with very high energy (Pian et al. 1998).

In Fig. (\ref{4}) we show the parameter space for Mrk 501 in the quiescent state. In 
this
state we can apply the relation for the Thomson case, given that the value of $\delta
$ required by the transparency condition is higher than $\delta _{KN}$.
We have used the following values for the observable quantities: $\nu _s=10^{16}$
Hz,
$\nu _c=10^{25}$ Hz, $\nu _sL_s(\nu _s) =10^{44}$ erg/s, $\nu _cL_c(\nu _c)=5\cdot 10^{43}$ erg/s,
$t_{var}\simeq 10$ h, $\alpha _1=0.5$ and $\alpha _2=1.75$.\\
For the quiescent state we have large uncertainties on the position of the Compton
peak: the boundaries of region A refer to values of $\nu _c\simeq 3\cdot
10^{24}$ Hz (upper line) to $\nu _c\simeq 3\cdot 10^{25}$ Hz.

Since a measurement of a time lag between variations in different frequencies
is lacking the associated constraint cannot be used.

The longer variability timescales and the lower luminosity with respect to the two
previous sources  allow in this case a
 value of $\delta $ around 10,  lower than those inferred in Mrk 421 and PKS 2155-304,
 as can be seen from Fig. (\ref{4}), while the value found for the magnetic
field intensity, $B\simeq 0.3$ is similar.
 
For the high state we must use the formulation in the KN limit: in fact the
Thomson condition (\ref{limite}) yields a value of $\delta \simeq 150$, highly
implausible. 
For the observables we have used the following values: $\nu _s=10^{19}$,
 $\nu _sL_s(\nu _s) =5 \cdot 10^{45}$ erg/s,
 $\nu _cL_c(\nu _c)=1.5 \cdot 10^{45}$ erg/s
and the variability time and the spectral indices as in the quiescent state.
The Compton peak frequency is not well determined since the spectra measured in the TeV 
range  suffer from various uncertainties (Samuelson et al. 1998). However all 
the measurements give
slopes steeper than 1 indicating that the peak is at lower energies. We therefore
consider values between $\nu_c=10^{25} - 10^{26}$ Hz.  

In the KN regime the parameter region allowed by the peak frequency ratio (region A)
has $B$ increasing with $\delta $ [see Fig. (\ref{5})]. 
The luminosity ratio (region B) has almost no intersection with region A for
values of $\delta $ allowed by the transparency condition. 
On the other hand the emitting region could be smaller than
imposed by the variability time scale by a factor larger than 3. In that case
region B would extend upwards defining a permitted region at $B > 1$, $\delta > 7-8$

However,  the condition of consistency for the value of $\gamma _b  \simeq 10^6$,
 given in the KN limit by eq. (\ref{sedicikn}) (region C), is definitely
outside the parameter region allowed by the other constraints,
 implying that the shape of the electron spectrum
cannot be determined balancing radiation losses and escape, unless again the
size of the region is much smaller than estimated from the  observed variability.
Since the flaring state lasts much longer than  the cooling time,
efficient continuous reacceleration processes are needed.

A comparison of the parameter space for the low and high states of Mkn 501 
indicates that not only $\gamma _b$ but also $B$ increased in the flaring state.

\section{Summary and Conclusions}

We have extended the analytic  treatment of the spectral constraints 
on the SSC homogeneous model for blazars initiated by GMD including KN corrections and
 adding the discussion
of other constraints now available.
The constraints can be summarized in the magnetic field-Doppler factor plane.
In principle the knowledge of $\nu_{s,peak}, \nu_{c,peak}, L_{s,peak}, L_{c,peak}$
plus $t_{var}$ allows the determination of B, $\delta$, $\gamma_b$, R and K.
The transparency condition, the time lag condition, the cooling-escape balance are
further restrictions that overconstrain the model. 

In practice large uncertainties on the observational quantities still exist, but
at least for the best observed  
 TeV blazar, Mrk 421, we can significantly constrain the parameter space.
 We find that
 the radiative interpretation of the lags is consistent
with the spectral constraints. Moreover, the break in the distribution is
consistent with the equilibrium between the energy losses and electron escape
timescale.
The derived value of the the Doppler factor is
moderately large ($\delta \simeq 25$), in agreement with the values found by specific
spectral fits (see e.g. Ghisellini \& Maraschi 1996; Mastichiadis \& Kirk 1997).

We have applied the analysis also to PKS 2155-304, a well observed object in the low
energy bands, detected in two occasions with EGRET in $\gamma $-rays. 
Assuming consistency between all the constraints (as found in the case of Mrk 421) 
we  estimate that the peak of the
Compton component should fall at about $10^{25}$ Hz: for higher values
 we do not find consistency between the spectral constraints and the
time lags as measured in the 1994 observational campaign (Urry et al. 1997).
 We  estimate the TeV
flux, which  could be measured by southern Cherenkov telescopes. 

The last object analized is Mrk 501, a TeV emitter that has shown a dramatic flare in
 April 1997. For the quiescent state we can apply the classical formulation, while
for the flaring state we must use the formulae in the KN limit.
The absence of information about time lags prevents to completely constrain
the model: however we can estimate that the value of the Doppler factor is about 10,
smaller than in the other two sources. The high state is consistent with a costant Doppler
factor, but impllies an increase in $\gamma_b$  and in the magnetic field. 
The required value of $\gamma_b$ is inconsistent with cooling-escape equilibrium, 
 indicating the need of an efficient continuous reacceleration mechanism.

In conclusion the analytic approach offers a solid basis for a general discussion
of the model and its applicability to  specific sources. It allows to derive
parameter ranges taking into account uncertainties in the "observable" quantities
and a better comprehension of the interrelations between various parameters.
We stress however that Table 2 has been derived applying a number of restrictive
choices and hypothesis for illustrative purposes and should not be taken at face value.

Individual fits remain essential to better understand the detailed shape
of the particle spectra and their origin. 

\noindent
{\it Note:} after the paper was accepted we learned about the detection of 
PKS 2155-304 by the Durham
Mark 6 Cerenkov Telescope (Chadwick et al. 1998) with a flux of $F(>0.3
TeV)=4.2\cdot10^{-11}$ ph cm$^{-2}$ s$^{-1}$, consistent with our estimate.

\appendix
\section{The Klein-Nishina limit}

In this Appendix we give the expression of the Compton emissivity for the electron
distribution assumed in this work and we prove eq. (\ref{intef}) for the frequency of the 
Compton peak in the KN limit. All the quantities are referred to the blob reference frame.\\
As previously assumed, we consider a population of relativistic electrons with a double
power-law distribution with indices $n_1$ and $n_2$ and break Lorentz factor $\gamma _b$:

\begin{equation}
N(\gamma )=\left\{ \begin{array}{ll}
                    K_1\gamma ^{-n_1}  &  \mbox{if $\gamma < \gamma _b$} \\
		    K_2\gamma ^{-n_2}  &  \mbox{if $\gamma _b < \gamma <\gamma _{max}$}
		   \end{array}
	   \right. 		
\label{distribution}
\end{equation}

\noindent
where $\gamma _{max} \gg \gamma _b$.\\
The coefficients $K_1$ and $K_2$ are related by the condition $K_1\gamma _b^{-n_1}=
K_2\gamma _b^{-n_2}$ which gives $K_2=K_1 \gamma _b^{n_2-n_1}$.\\
The seed photons are those emitted by the synchrotron mechanism and in the blob
reference frame the radiation energy density can be written as (hereafter we use the
parameter $x^{\prime}=h\nu ^{\prime}/mc^2)$

\begin{equation}
\epsilon (x^{\prime}_0) =\left\{ \begin{array}{ll}
             \epsilon _0 \left( \frac{x^{\prime}_0}{x^{\prime}_s} \right)^{-\alpha _1} & \mbox{for 
             $x^{\prime}_{min}<x^{\prime}_0<x^{\prime}_s$} \\
             \epsilon _0 \left( \frac{x^{\prime}_0}{x^{\prime}_s} \right)^{-\alpha _2} & \mbox{for 
             $x^{\prime}_{max} > x^{\prime}_0 > x^{\prime}_s$}
                   \end{array}
            \right.
\label{endens}
\end{equation}

\noindent
where $\alpha _i=(n_i-1)/2$.\\
Using a $\delta $-approximation (see e.g. Coppi \& Blandford 1990) 
the Compton emissivity in the blob reference frame can be written as:

\begin{equation}
j(x^{\prime})=\frac {\sigma _Tmc^3}{8\pi h} \int _{x_1} ^{x_2} N \left[ \left( \frac{3x^{\prime}}
{4x^{\prime}_0} \right)^{1/2} \right] \left( \frac {3x^{\prime}}{4x^{\prime}_0} \right)^{1/2} 
\frac{\epsilon (x^{\prime}_0)}{x^{\prime}_0} dx^{\prime}_0
\end{equation}

\noindent
where the limits of the integral are analyzed in the following.\\
For $N(\gamma )$ and $\epsilon (x_0)$ given above we can split the integral in four
similar integrals:

\begin{equation}
j(x^{\prime})=I_{1,1}+I_{1,2}+I_{2,1}+I_{2,2}
\label{sum}
\end{equation}

\noindent
The first term is:

\begin{equation}
I_{1,1}=C(\alpha _1)K_1\epsilon _0\left( \frac {x^{\prime}}{x^{\prime}_s} \right)^{-\alpha _1}
\ln \frac{x_1}{x_2}
\end{equation}

\noindent
with $C(\alpha _1)=(4/3)^{-\alpha _1}(\sigma _Tmc^3)/(8\pi h)$.
The limits are:

\begin{equation}
x_1=\max \left[x^{\prime}_{min}, \frac{3x^{\prime}}{4\gamma _b^2} \right]
; \  x_2=\min \left[x^{\prime}_s, \frac{3x^{\prime}}{4}, \frac{3}{4x^{\prime}} \right] 
\end{equation}

\noindent
The term $3/4x^{\prime }$ in the upper limit express the KN
suppression of the high energy emission. It originate from the step approximation of
the KN cross-section, that can be rewritten as the condition that the scattering are
in the Thomson regime up to:
\begin{equation}
\gamma x^{\prime }_0< \frac{3}{4}
\end{equation}
\noindent
This condition, with the use of 
\begin{equation}
x^{\prime }=(4/3)\gamma ^2x^{\prime }_0
\end{equation}
\noindent
gives:
\begin{equation}
x^{\prime }_0 < \frac{3}{4} \frac {1}{x^{\prime }}
\end{equation}  
\noindent
that is the condition used in the integrals.

\noindent
The second integral is:

\begin{equation}
I_{1,2}=C(\alpha _1)K_1\epsilon _0x^{\prime -\alpha _1}x_s^{\prime \alpha _2} \frac{1}{\alpha
_2-\alpha _1} \left[ x_1^{\alpha _1-\alpha _2} - x_2^{\alpha _1-\alpha _2}
\right]
\end{equation} 

\noindent
where:

\begin{equation}
x_1=\max \left[ x^{\prime}_s, \frac{3x^{\prime}}{4\gamma _b^2} \right] ; \      
x_2=\min \left[\frac{3x^{\prime}}{4}, \frac{3}{4x^{\prime}} \right]
\end{equation}

The third term can be written as:

\begin{equation}
I_{2,1}=C(\alpha _2)K_1\gamma _b^{2(\alpha _2-\alpha _1)} \epsilon _0x^{\prime -\alpha _2}
x_s^{\prime \alpha _1} \frac{1}{\alpha
_2-\alpha _1} \left[ x_2^{\alpha _2-\alpha _1} - x_1^{\alpha _2-\alpha _1}
\right]
\end{equation} 

\noindent
where, in this case $C(\alpha _2)=(4/3)^{-\alpha _2}(\sigma _Tmc^3)/(8\pi h)$ and:

\begin{equation}
x_1=\max \left[x^{\prime}_{min}, \frac {3x^{\prime}}{4\gamma _{max}^2} \right]; \  
x_2=\min \left[ x^{\prime}_s, \frac {3x^{\prime}}{4\gamma _b^2}, \frac{3}
{4x^{\prime}}  \right]
\end{equation}

The last term is:

\begin{equation}
I_{2,2}=C(\alpha _2)K_1\gamma _b^{2(\alpha _2-\alpha _1)} \epsilon _0 \left( \frac{x^{\prime}}
{x^{\prime}_s}
\right)^{-\alpha _2} \ln \frac{x_2}{x_1}
\end{equation}

with:
\begin{equation}
x_1=\max\left[ x^{\prime}_s,\frac {3x^{\prime}}{4\gamma _{max}^2} \right]; \  
x_2=\min \left[ x^{\prime}_{max},\frac {3x^{\prime}}{4\gamma _b^2},\frac{3}{4x^{\prime}}  \right]
\end{equation}

The Compton emissivity around the peak in the KN limit is dominated by the contribution
of $I_{1,1}$ and $I_{2,1}$ (see Fig. 2 d). One can easily see that the emissivity can 
be written as:

\begin{equation}
j(x^{\prime})=C(\alpha _1)K_1\epsilon _0\left( \frac {x^{\prime}}{x_s^{\prime}} \right)^{-\alpha _1}
\ln \left[ \gamma _b^2 \frac{1}{x^{\prime 2}} \right] +
C(\alpha _2)K_1 \epsilon _0x^{\prime -\alpha _1}
x_s^{\prime \alpha _1} \left( \frac {3}{4} \right)^{\alpha _2-\alpha _1} 
\frac{1}{\alpha _2-\alpha _1}
\end{equation}
\noindent
>From this expression we can find the frequency of the Compton peak, that is the maximum
of the function $j(x^{\prime})x^{\prime} $. Taking the derivative we can write the maximum condition as:

\begin{equation}
\ln \left[ \gamma _b^2 \frac{1}{x^{\prime 2}} \right] + 
\frac{1}{\alpha _2-\alpha _1} - \frac {2}{1-\alpha _1}=0
\end{equation}

\noindent
which gives the value of the frequency of the Compton peak:

\begin{equation}
x^{\prime}_c=\gamma _b \exp \left[ \frac{1}{\alpha _1-1} + 
\frac{1}{2(\alpha _2-\alpha _1)} \right] 
\end{equation}

\section{Constants used in the text}

The constant $\xi (\alpha _1, \alpha _2)$ has the expression:

\begin{equation}
\xi (\alpha _1, \alpha _2) = 2\cdot10^{-7} \left( \frac{h}{mc^2} \right)^{\frac{4\alpha
_1-3}{1-\alpha_1} } \left[ \frac{f(\alpha _1, \alpha _2)}{5.7\cdot 10^{39}}
\right] ^{1/(1- \alpha _1)} g(\alpha _1, \alpha _2)^{ \frac{2-3\alpha _1}{1-\alpha _1} }
\end{equation}

\noindent
In Table 3 we give the value of $\xi (\alpha _1, \alpha _2)$ for some common values of
the spectral indices.

\clearpage

\vskip 1.5 truecm

\centerline{ \bf Figure Captions}

\vskip 1 truecm

\figcaption[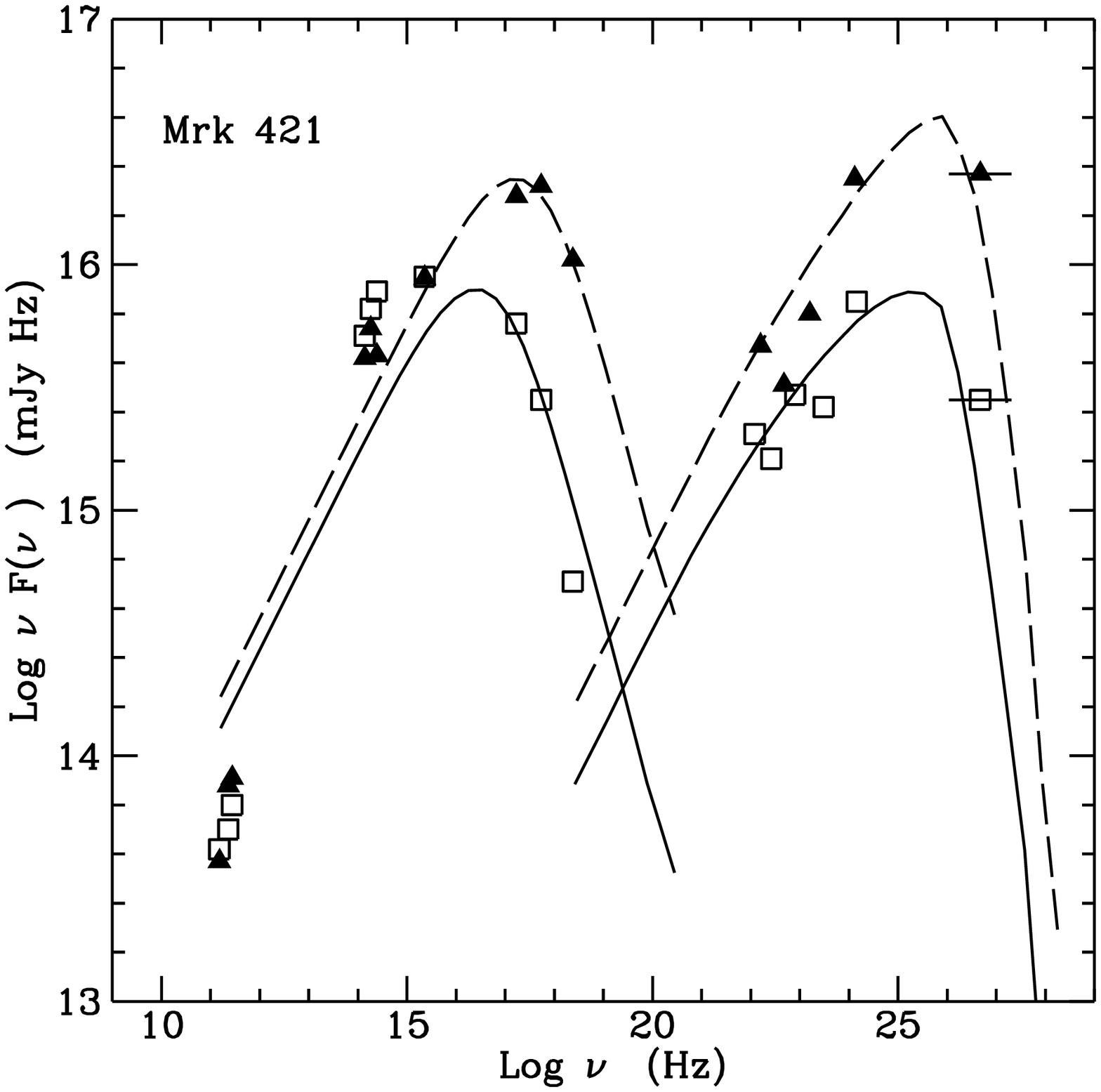]{ SED of Mrk 421 in two epochs (triangles and squares, data from Macomb et al.
1995). Spectra calculated with the homogeneous
SSC model with the electron energy distribution $N(\gamma )=K\gamma ^{-n_1}(1+\gamma / \gamma
_b)^{n_1-n_2}$ are superposed to the data: both peaks are reasonably well described by 
this
 double power-law approximation. The models are calculated with the following
parameters: low state: $R=2.7\cdot 10^{15} cm$, $B=0.15 G$, $K=1.7\cdot 10^5$,
$\delta=25$, $\gamma _b=5.6\cdot 10^4$, $n_1=2.2$, $n_2=4.5$; flaring state:  as in
the low state but $\gamma _b=1.4\cdot 10^5$, $K=2.4\cdot 10^5$. Note that in the radio
band ($\nu < 10^{11}$ Hz) the emission is selfabsorbed. \label{prima} }

\figcaption[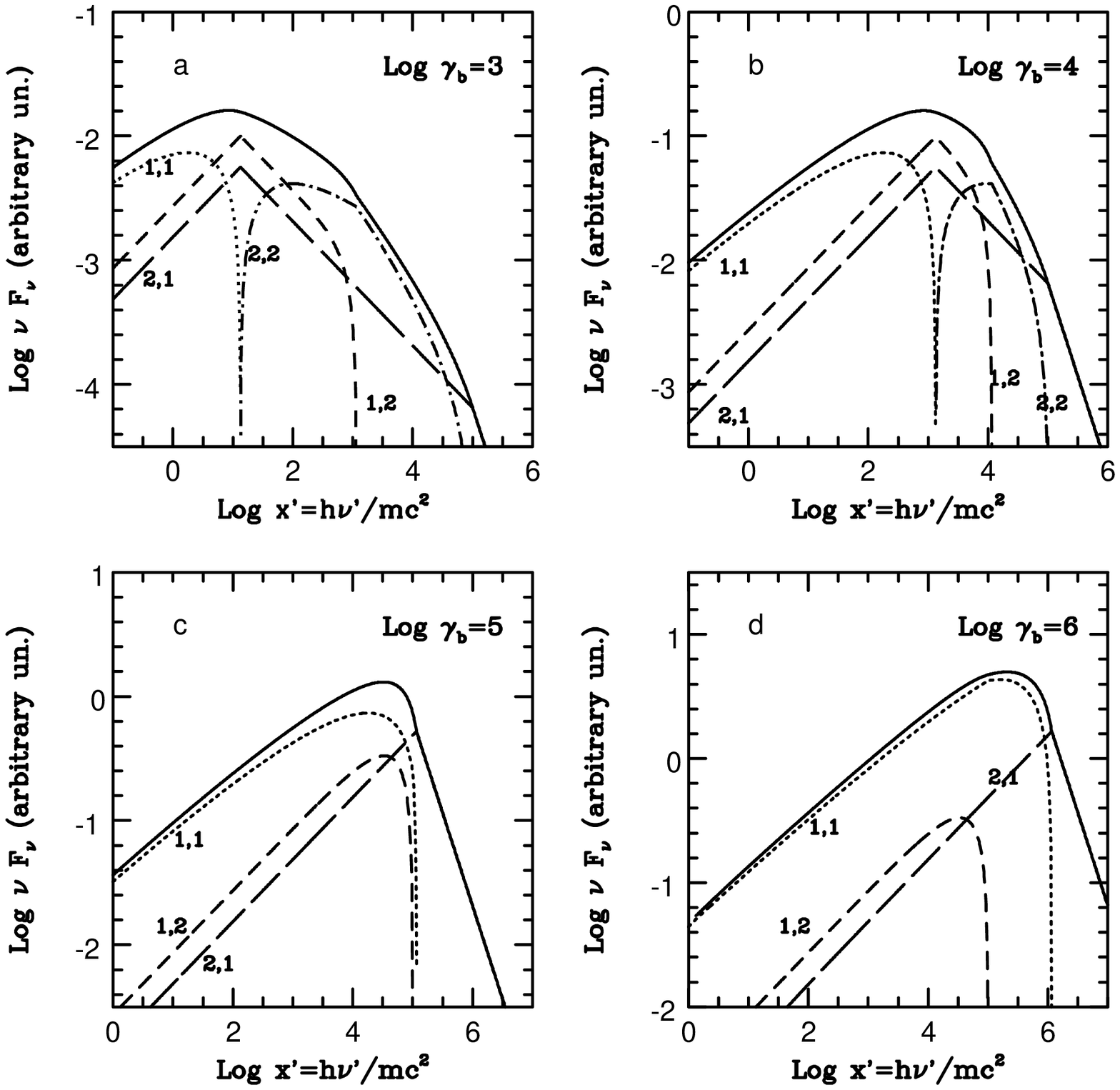]{ Inverse Compton SSC spectra calculated using the formulae described in the
Appendix (the frequency is expressed as $x^{\prime }=h\nu ^{\prime }/mc^2$). 
The solid line is the total 
spectrum, while the other curves show the contributions of electrons and photons with
different energies (see the following). The spectra are obtained for fixed
synchrotron peak frequency ($\nu _s^{\prime }=10^{15} Hz$) and spectral indices ($\alpha
_1=0.5$, $\alpha _2=1.5$), but for different Lorentz factors of the electrons at the
break. The dotted line shows the contribution from the electrons and the photons
before the break ($I_{1,1}$), the short dashed line shows the
contribution from the $I_{1,2}$ branch (low energy electrons and high energy photons),
the long dashed line is the contribution
from $I_{2,1}$ (high energy electrons and low energy photons) and finally 
the dot-dashed curve is the $I_{2,2}$
spectrum (the contribution from the high energy photons and high energy electrons).
 In panel a) $\gamma _b=10^3$ and the Compton peak is
produced in the Thomson limit: all the contributions are present and the peak frequency
is given by $\nu _c=(4/3)\gamma _b^2\nu _s$. For higher values of $\gamma _b$ [panel
b)]
the contribution from $I_{2,2}$ begins to be suppressed; when the peak is in the KN 
regime
[panel c) and d)] the latter contribution vanishes, while also the 
contribution from $I_{1,2}$ is strongly suppressed. In the strong KN limit, only 
the $I_{1,1}$ and
 $I_{2,1}$ branches are important. Note that the spectrum after the peak in
the KN limit is a power-law with the same index that the $I_{2,1}$ component, $\alpha
_{KN}=2\alpha _2-\alpha _1$ \label{trans} }

\figcaption[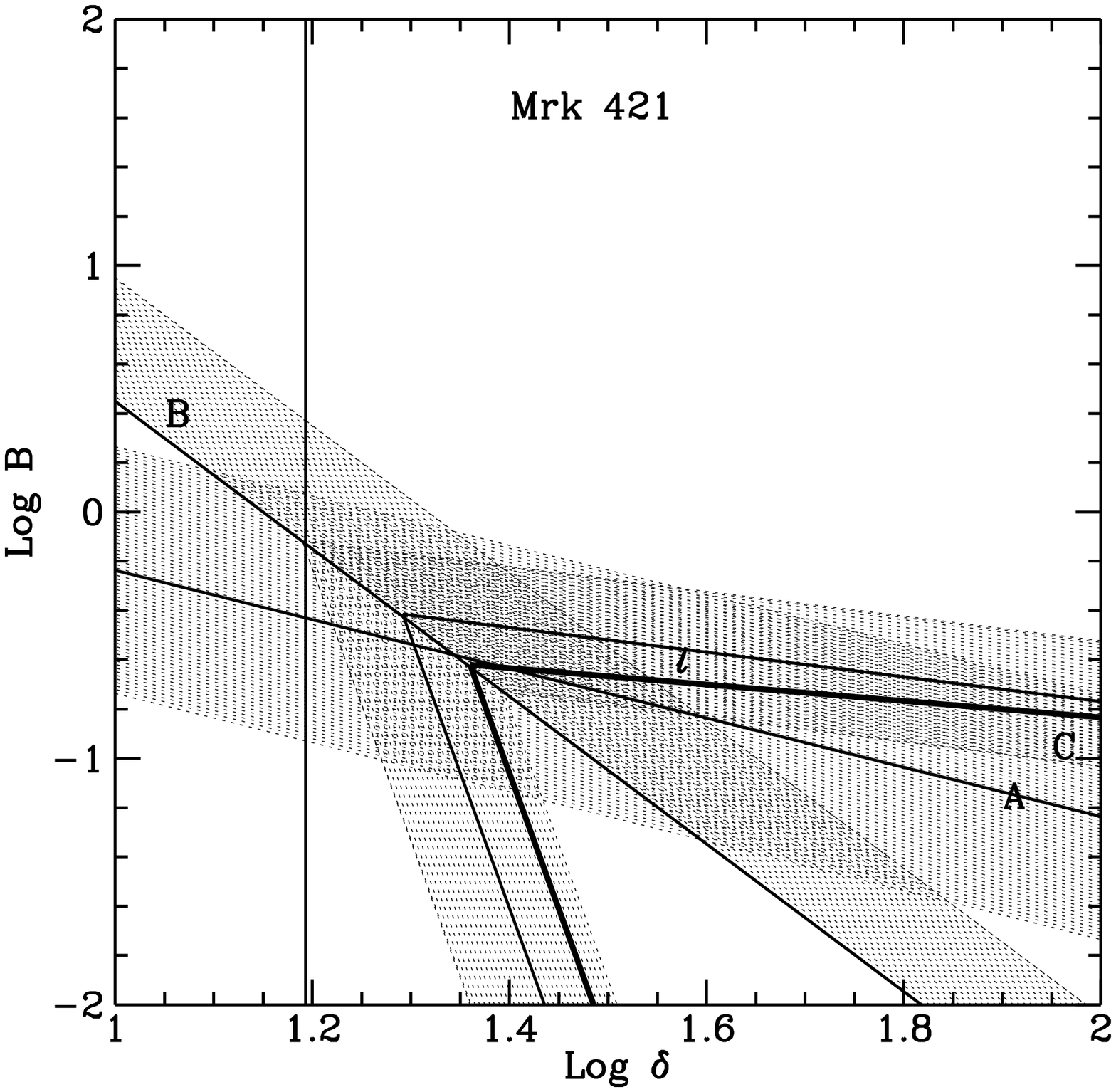]{ Constraints on the parameter space for Mrk 421. Region A 
represents
the peak frequency constraint. The region takes into account an uncertainty of a factor
3 in the position of the Compton peak, the upper line is calculated for $\nu
_c=2\cdot 10^{25}$ Hz, the lower line for $\nu _c=1.8\cdot10^{26}$ Hz and the
intermediate line for $\nu _c=6\cdot10^{25}$ Hz. Region B
is obtained from the peak luminosity constraint and takes into account an uncertainty
of a factor 3 in the dimension of the region (the upper line is calculated for
$t^{\prime }_{var}=1/3t_{var}$). The region C is obtained from the equilibrium of
cooling and escape for electrons with Lorentz factor $\gamma _b$; the region is plotted taking into account an uncertainty of
a factor 3 in the value of $t_{var}$ and in the value of $\beta _{esc}$ (1-1/3). 
The thick line {\it l} is obtained from the abserved time lag between
variations in different X-ray band. The vertical line shows the lower limit on $\delta
$ given by the transparency condition. The plot shows that consistency of all the
constraints is achieved for $\delta \simeq 25$ and $B \simeq 0.2$ \label{2} }

\figcaption[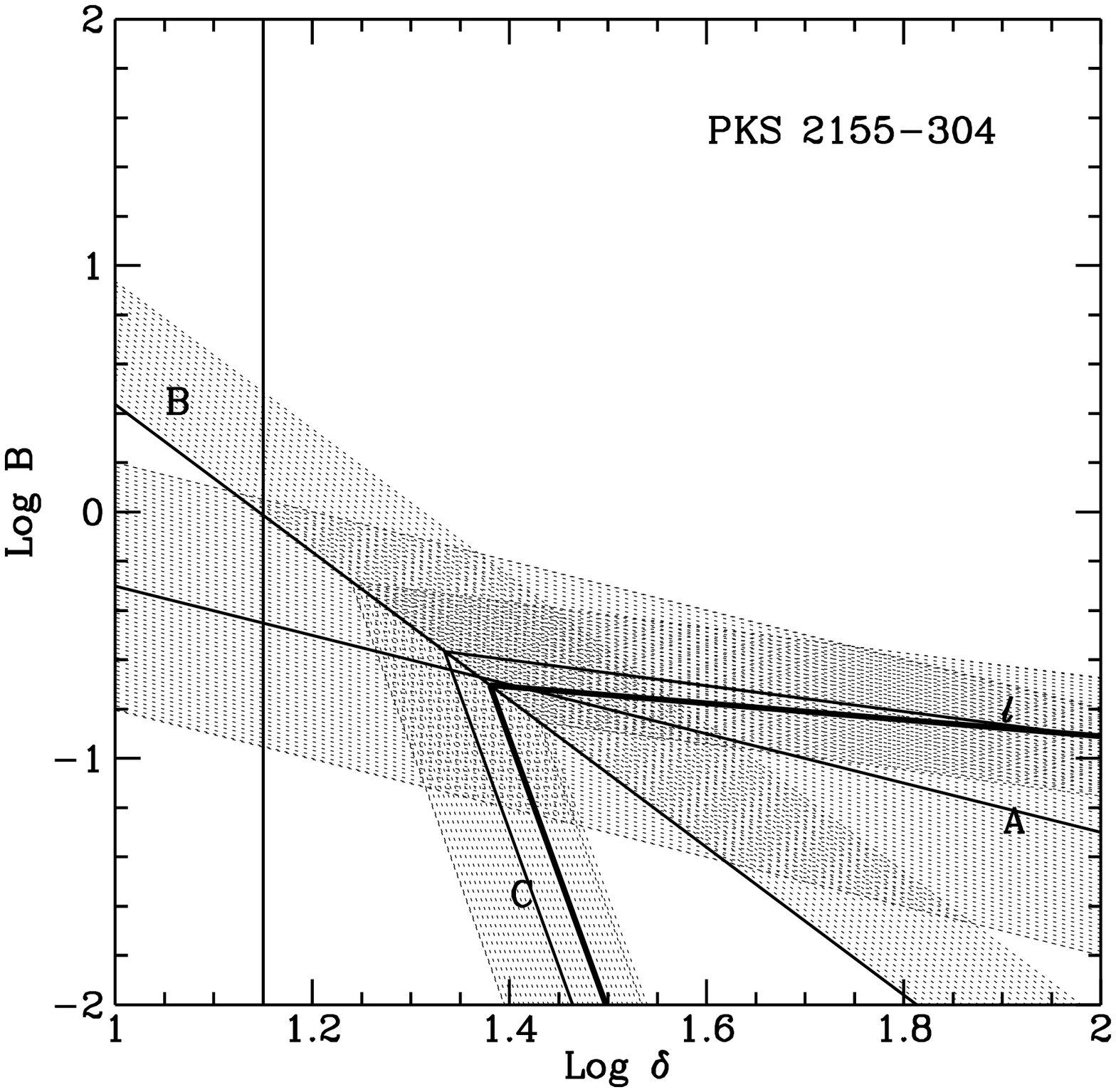]{ Parameter space for PKS 2155-304. Lines and regions are 
labelled as in Fig. 3. Region A takes into account an uncertainty of a
factor $3$ in the Compton peak frequency ($2.5\cdot 10^{24}$--$2.5\cdot 10^{25}$).
\label{3} }

\figcaption[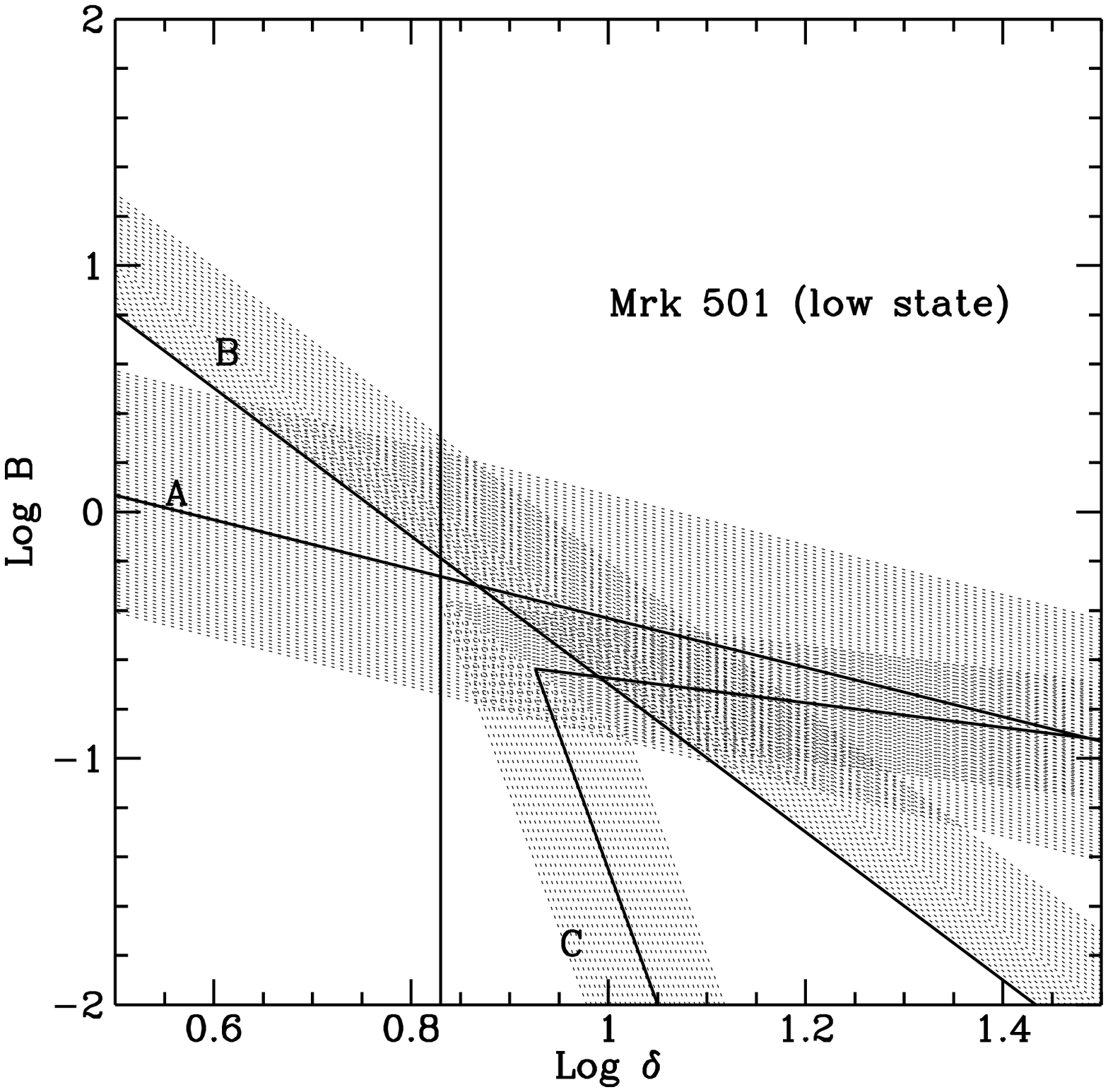]{ Parameter space for Mrk 501 (low state). For this object we do not have
information about delays in the variability and therefore in the diagram we cannot
plot the line {\it l}. \label{4} }

\figcaption[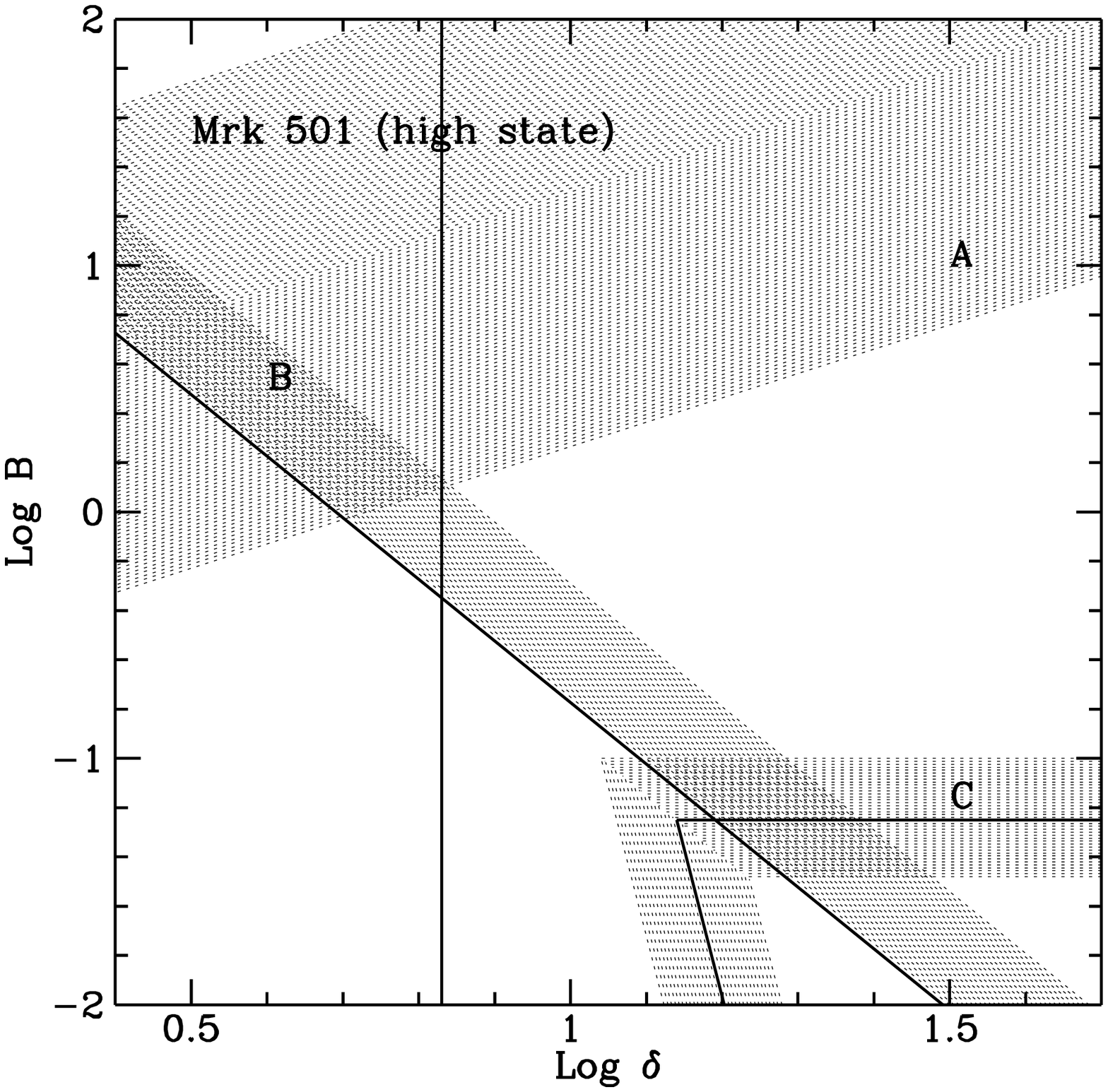]{ Parameter space for Mrk 501 (high state). In this case we have used the
formulae for the KN limit and region A has a positive slope. Note that the break
energy--balance
condition between energy losses and escape  
(region C) is not consistent with the other constraints.\label{5} }

\clearpage
\begin{deluxetable}{lcrrrr}
\tablecaption{Input observational quantities}
\tablehead{
\colhead{ } & \colhead{ } & \colhead{Mrk 421}  & \colhead{PKS 2155-304}      & \colhead{Mrk 501
(l.s.)} & \colhead{Mrk 501 (h.s.)}  } 

\startdata
$\nu _s$ (Hz) && $3\cdot10^{16}$ & $10^{16}$ & $10^{16}$ & $10^{19}$ \\ 
$\nu _c $ (Hz) && $2\cdot10^{25}-2\cdot10^{26}$ & $2.5\cdot10^{24}-2.5\cdot10^{25}$ & $3\cdot10^{24}-3\cdot10^{25}$ & $10^{25}-10^{26}$\\ 
$\nu _sL_s(\nu _s)$ && $4.7\cdot 10^{44}$ & $1.5\cdot 10^{46}$ & $10^{44}$ & $5\cdot 10^{45}$\\
$\nu _cL_c(\nu _c)$ && $6.3\cdot 10^{44}$ & $1.5\cdot 10^{46}$ & $5\cdot 10^{43}$&$1.5\cdot 10^{45}$ \\ 
$t_{var}$ (h) && $1 $ & $5.5$ & $10 $& $10 $\\ 
$\tau _{obs}^a(s)$ && $4200$ & 4700& - & -\\
\tablenotetext{a}{lag between photons of 5 keV and 0.75 keV}
\enddata 
\end{deluxetable}

\clearpage
\begin{deluxetable}{lrrrr}
\tablecaption{Physical parameters estimated from the present analysis}
\tablehead{
\colhead{ } & \colhead{Mrk 421}  & \colhead{PKS 2155-304}      & \colhead{Mrk 501 (l.s.)} & 
\colhead{Mrk 501 (h.s.)}  }

\startdata
$\delta $ &$22-35$ & $24-38$ &$8-20$ & $7$\\
$B$ (G) & $0.3-0.2$ & $0.3-0.2$ & $0.5-0.1$ & $1$\\
$\gamma _b^a$ & $4-3$& $2-1.8$ & $3-4$ & $60$\\
$R^b$ & $3-1$ & $15-7$ & $9-7$ & $2.5$\\
\tablenotetext{a}{in units of $10^4$. }
\tablenotetext{b}{in units of $10^{15}$ cm.}
\enddata
\end{deluxetable}

\clearpage
\begin{deluxetable}{lrr}
\tablewidth{25pc}
\tablecaption{Values of $\xi (\alpha _1, \alpha _2)$ for some values of the spectral
indices }
\tablehead{
\colhead{$\alpha _1$} &  \colhead{$\alpha _2$} & \colhead{$\xi (\alpha _1, \alpha _2)$} } 

\startdata
0.5 & 1.25 & $9.13\cdot 10^{-46}$\\
0.5 & 1.75 & $2.17\cdot 10^{-46}$\\
0.75 & 1.25 & $1.54\cdot 10^{-161}$\\
0.75 & 1.75 & $1.1\cdot 10^{-165}$
\enddata
\end{deluxetable}

\clearpage

\begin{figure}
\centerline{\plotone{intro.ps}}
\end{figure}

\clearpage

\begin{figure}
\centerline{\plotone{quattro.ps}}
\end{figure}

\clearpage

\begin{figure}
\centerline{\plotone{421n.ps}}
\end{figure}

\clearpage

\begin{figure}
\centerline{\plotone{2155n.ps}}
\end{figure}

\clearpage

\begin{figure}
\centerline{\plotone{501n.ps}}
\end{figure}

\clearpage

\begin{figure}
\centerline{\plotone{501hn.ps}}
\end{figure}

\end{document}